\documentclass[showpacs,twocolumn,floatfix,superscriptaddress]{revtex4}
\usepackage{graphicx,float,array,afterpage,ltxtable,}
\usepackage[english]{babel}
\usepackage{amssymb,amsmath,indentfirst}
\usepackage[T1]{fontenc}
\usepackage{epsfig}
\usepackage{rotating}

\begin{document}

\title{Experimental observation of a complex periodic window}
\author{D. M. Maranh\~ao}
\affiliation{Instituto de F\'isica, Universidade de S\~ao Paulo, Caixa Postal 66318,
05315-970, S\~ao Paulo, SP, Brazil}
\author{M. S. Baptista}
\affiliation{Max-Planck Institute f\"ur Physik komplexer Systeme, N\"othnitzerstr. 38,
D-01187 Dresden, Germany}
\author{J. C. Sartorelli}
\author{I. L. Caldas}
\affiliation{Instituto de F\'isica, Universidade de S\~ao Paulo, Caixa Postal 66318,
05315-970, S\~ao Paulo, SP, Brasil}

\begin{abstract}
The existence of a special periodic window in the two-dimensional parameter space of an 
experimental Chua's circuit is reported. One of the main reasons that makes 
such a window special is that the observation of one implies that other 
similar periodic windows must exist for other parameter values. However, 
such a window has never been experimentally observed, since its size in 
parameter space decreases exponentially with the period of the periodic 
attractor. This property imposes clear limitations for its experimental 
detection.
\end{abstract}

\maketitle



The emergence of regular behavior is one of the most studied topics in
nonlinear dynamical systems. It is known that by the changing of an accessible
parameter, chaos \cite{jacobson:1981} and periodic \cite{graczyk:1997}
behaviors will be observed. 

The expectation of finding stable periodic behavior inside chaotic regions in
parameter space depends on the sizes and shapes of the parameter regions,
regarded as \textit{periodic windows} (PWs), for which stable periodic orbits
(POs) are found. A PW is a region in parameter space that indicates parameter
values for which one finds the lowest periodic attractor of period $P$, plus
the period-doubling cascade with attractors of period $P^{2n}$, with $n
\in \mathcal{N}$ \cite{comment_PW}.

For systems whose chaotic attractors have only one positive Lyapunov exponent
as the Chua's circuit, considered in this experiment, a special type of PW,
regarded as \textit{complex periodic windows} (CPWs), is everywhere observed in
parameter space. The appearance of one such window implies in the appearance
of an infinite number of self-similar others that appear side by side aligned
along a direction. In addition, CPWs have an extended characteristic in the
parameter space. They visit large portions of the parameter space; i.e. one
can still stay in the same periodic windows even if especially large variations
in two control parameters are made. Due to these two characteristics an
arbitrary change in only one accessible parameter can replace chaos by
periodic behavior, or vice versa. So a better understanding of a CPW is
relevant to applications that relay either on a robust periodic oscillation,
as mechanical machines, or on a robust chaotic system, as chaos-based
communication \cite{chaos_based}.

These CPWs, regarded as \textit{shrimps} \cite{gallas}, were extensively
studied in maps \cite{hunt:1997,barreto:1997} and in periodically forced maps
\cite{baptista_CHAOS1996,murilo1}. However, only recently were these windows 
numerically observed in systems of ordinary differential equations
\cite{gallas:2005,baptista:1997}. The reason is that the parameter interval
length, $\Delta {\mathcal{P}}$, of a CPW scales exponentially with $-P$, where
$P$ is the period $P$ of the lowest-period periodic attractor of the CPW
\cite{hunt:1997}.  Since CPWs have usually higher $P$, they are too tiny to be observed, even
though these tiny windows are extended in parameter space.

This exponential scaling clearly imposes limitations on the
experimental detection of such a periodic window, and arguably due to that, 
they have never been experimentally reported. However, for the Chua's
circuit, it was numerically shown in Ref. \cite{baptista:1997} that such
CPWs possessing a low value for the lowest-period periodic attractor ($P$=4)
exist. This work is dedicated to experimentally report, for the first time,
such a CPW.

To certify that we observed a CPW, we show that there exists curves in
parameter space where the POs are super-stable, and that these curves cross
transversally at least twice, a necessary condition that defines a CPW.  These
parameter curves are detected by the indirect method of noting the parameter
values at which the symbolic sequences, encoding the type of POs existing
within the CPW, change.

\begin{figure}[htb]
\begin{center}
\includegraphics[width=9.0cm,height=10cm,viewport=0 20 550
630]{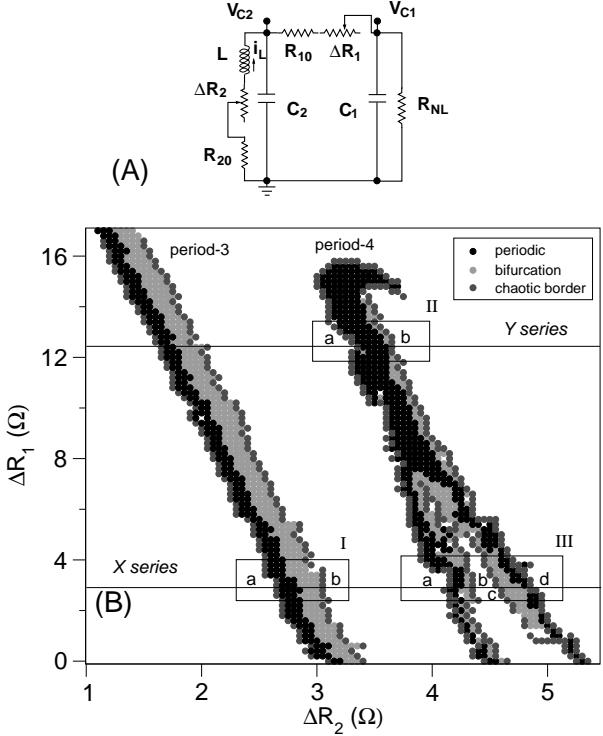}
\end{center}
\caption{(A) Scheme of Chua's circuit. Their component values are: 
$R_{10}\approx 1.4 k\Omega$, $R_{20}\approx 37 \Omega$, $C_1\approx 4.7$nF, 
$C_2\approx 56$nF, $L\approx 9.2$mH. (B) Parameter space for the experimental
Chua's circuit showing the period-3 and period-4 windows. Filled black
circles represent parameters for which the lowest period POs are observed,
filled light gray circles represent the higher period PO that appear by
periodic-doubling bifurcations, and filled dark gray circles represent
parameters for the closest chaotic attractor to the PWs. The straight lines
indicate parameter values for the sets of time series \emph{X} and \emph{Y}.}
\label{fig1}
\end{figure}

The well known Chua's circuit is shown in Fig.~\ref{fig1}(A). The control
parameters are $R_{1}=R_{10}-\Delta R_{1}$ and $R_{2}=R_{20}-\Delta R_{2},$
where $R_{10}$ and $R_{20}$ have fixed values, $\Delta R_{1}$ and $\Delta
R_{2}$ are varied by precision potentiometers, with steps of $50m\Omega $ and
$200m\Omega $, in the ranges $[0,17]\Omega$ and $[1,$ $5.5$]$\Omega$,
respectively. We obtained time series by recording the $V_{C_{1}}(t)$ voltage
with a 12-bit analog-to-digital converter (ADC) at the rate of
400~ksamples/s. All the attractors were reconstructed by the Takens
method~\cite{takens} with time-delay $\tau =45.0~\mu $s, which corresponds to
18 data points. Then, the reconstructed attractors are made discrete by
measuring $V_{C1}(t+\tau)$ when the reconstructed trajectory reaches the
section $V_{C1}(t)$=-2.25V in a clockwise orientation. The value of
$V_{C1}(t+\tau)$ when the reconstructed trajectory realizes its $n$th
crossing in this section is denoted by $V_{C1}^n$.

In Fig. \ref{fig1}(B), we show the parameter space of this circuit. There,
solid black circles represent parameter values for which one obtains the
lowest period PO. Along the left border between chaos and the PW [parameters
indicated by letters "a" within the boxes of Fig \ref{fig1}(B)] in these two
PWs, chaos is replaced by a stable (period-3 or period-4) attractor by a
tangent bifurcation by increasing $\Delta R_2$. In the other borders,
[parameters indicated by letters "b", "c" and "d"], the lowest-period PO
inside the PWs bifurcate and chaos (outside the PW) is reached after a
period-doubling cascade by modifying $\Delta R_2$. 

To illustrate our analysis techniques, we first use a symbolic representation
to characterize the lower period POs that appear for the parameters nearby the
borders between the PWs and the chaotic regions, indicated by the letters
\textbf{a},
\textbf{b}, \textbf{c} and \textbf{d}, in the boxes I, II, and III, in 
Fig~\ref{fig1}(B). We use data sets collected varying $\Delta R_{2}$ along the
lines $X$ and $Y$ for $\Delta R_{1}\mathit{=3.0}\Omega$ and $\Delta
R_{1}$\textit{=}$\mathit{12.5}\Omega$, respectively, as shown in
Fig.~\ref{fig1}(B).

\begin{figure}[tbh]
\begin{center}
\includegraphics*[width=9.0cm,height=10cm,viewport=0 0 550 720]{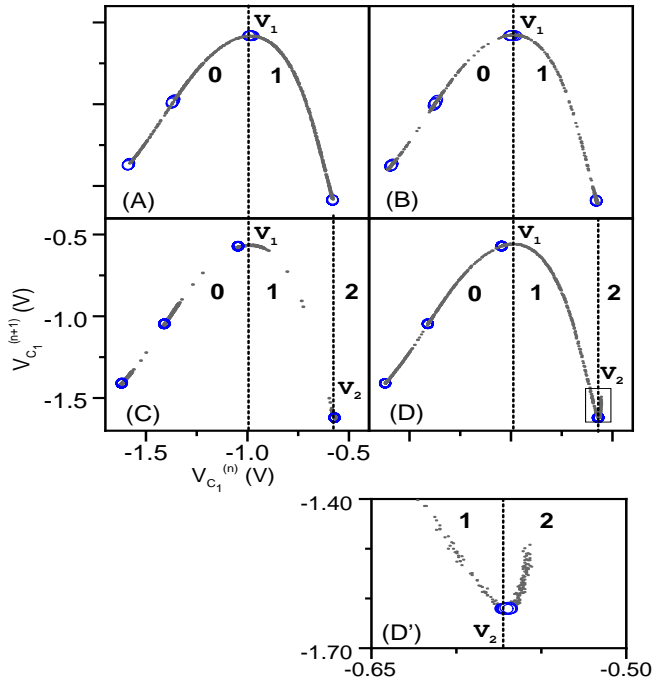}
\end{center}
\caption{[Color online] Return maps (black points) of the Poincar\'{e}
section of the chaotic attractors obtained using the parameters indicated by
the borders \textbf{a}, \textbf{b},
\textbf{c} and \textbf{d} in box III of Fig. \protect\ref{fig1}, respectively
Figs. (A)-(D). We also show the return maps (blue circles) of the periodic
attractors obtained for the closest parameters to these borders. The vertical
lines, passing through the maximum and the minimum define the partition
points. In (D') is shown a zoomed view of the minimum of the return map in
(D).}
\label{fig5}
\end{figure}

The symbolic characterization of these POs is done by encoding them by the
approach in Ref. \cite{macalum}, using the properties of the nearby chaotic
attractors. The return maps of the reconstructed chaotic attractors for
parameters in the borders \textbf{a}, \textbf{b}, \textbf{c}, and \textbf{d}
in box III, are shown in Figs. \ref{fig5}(A-D). These maps as well as the
other chaotic attractors at the borders in both period-3 and period-4 windows
display return maps typical of either unimodal (one maximum) or bi-modal (one
maximum and one minimum) maps, and they can be partitioned by the critical
points. The partitions are in the maximal and minimal points, assigned by
$V_{1}$ and $V_{2}$. So a trajectory point in the interval $V_{C1}<V_{1}$
is encoded by '0', a trajectory point in the interval $V_{1}<V_{C1}<V_{2}$
is encoded by '1', and a point in the interval $V_{C1}>V_{2}$ is encoded by
'2'. A stable period-$P$ orbit can be encoded by comparing its mapping with
the mapping of the nearby chaotic attractor, and depending on the position of
the POs points with respect to the partition points, a PO can be encoded by a
sequence ${s_{1}s_{2}\ldots s_{P}}$, where $s_{i}$ is a symbol of the
alphabet $s_{i}=\{0,1,2\}$. For chaotic attractors close to the borders with
the period-3 window, in box I, the chaotic returning maps are uni-modal,
with only one critical point $V_{1}$.  In the side \textbf{a} of the
window, in box I, we obtain the symbolic sequence ${101}$ and in the right
side \textbf{b}, the sequence ${100}$. All the POs in the left side of this
window are encoded by 101 and the ones on the right side by 100. The period-4
POs, in the period-4 window, close to the borders \textbf{a}, \textbf{b},
\textbf{c} and \textbf{d}, in box III, [whose return maps can be seen in
Figs. \ref{fig5}(A-D), respectively] are encoded by the sequences ${1001}$,
${1000}$, ${2000}$ and ${2000}$, respectively.

In fact, as one varies a control parameter, the symbolic sequence of a
stable PO changes if some periodic point crosses a critical point of the
return map \cite{metropolis}. This mechanism is responsible for the changes
in the symbolic sequences of the stable POs in the period-3 window. There,
the symbolic sequence ${101}$ changes to ${100}$ when the PO crosses the
critical point $V_{1}$.

We name $\xi $ the return map of the closest chaotic attractor to the period-%
$P$ PO, and $\mathcal{O}$ a stable PO with points ${V_{C_{1}}^{1},\ldots
,V_{C_{1}}^{P}}$. Assuming that the return map $\xi $ can be used as an
approximation to calculate the first derivative of the orbit points of a PO
inside a PW, then the orbit $\mathcal{O}$ is stable if 
\begin{equation}
\Delta <1  \label{estabilidade}
\end{equation}
with $\Delta =|\prod_{i=1}^{P}\frac{d\xi }{dV_{C_{1}}^{i}}|$. If a PO
contains a critical point, a point on the extremum of the map, $\Delta =0$, 
and we say such an orbit is superstable. 
For parameters $\epsilon $-close to a parameter for which a super-stable PO
exists, Eq. (\ref{estabilidade}) is satisfied, which means that it exists a
PW in the neighborhood of parameter lines for which $V_{C_{1}}^{i}=V_{1}$.

A similar mechanism governs the changes in the symbolic sequences of the
stable POs inside the period-4 region. The difference now is that we have two
critical points, $V_{1}$ and $V_{2}$, which makes Eq. (\ref{estabilidade}) 
satisfied in parameter curves for which either $V_{C_{1}}^{i}=V_{1}$ (which
defines the critical curve $S_{V1}$) or $V_{C_{1}}^{i}=V_{2}$ (which defines
the critical curve $S_{V2}$), or $V_{C_{1}}^{i}=V_{1}$ and
$V_{C_{1}}^{i}=V_{2}$. It is typical for this type of CPW that the PW appears
not only for the parameter point for which $V_{C_{1}}^{i}=V_{1}$ and
$V_{C_{1}}^{i}=V_{2}$, a zero measure point in parameter space, but also along
the curves $S_{V1}$ or $S_{V2}$. These two curves form the spines introduced
in Refs. \cite{barreto:1997,murilo1}.

Three important characteristics grant to this window the status of being a
CPW: (i) if there is one CPW, then a countable infinite number of others must
exist, with sizes that decreases exponentially [Eq. (\ref{rene})] as the
period of the POs increase. (ii) The two critical curves $S_{V1}$ and $S_{V2}$
cross transversally at least twice. For the parameters where the crossings
happen, the PO has an orbit point $V_{C_{1}}=V_{1}$ and another
$V_{C_{1}}=V_{2}$. (iii) POs with the same period coexist.


\begin{figure}[tbh]
\begin{center}
\includegraphics*[width=8.0cm, height=10.5cm,viewport= 0 0 500 700]{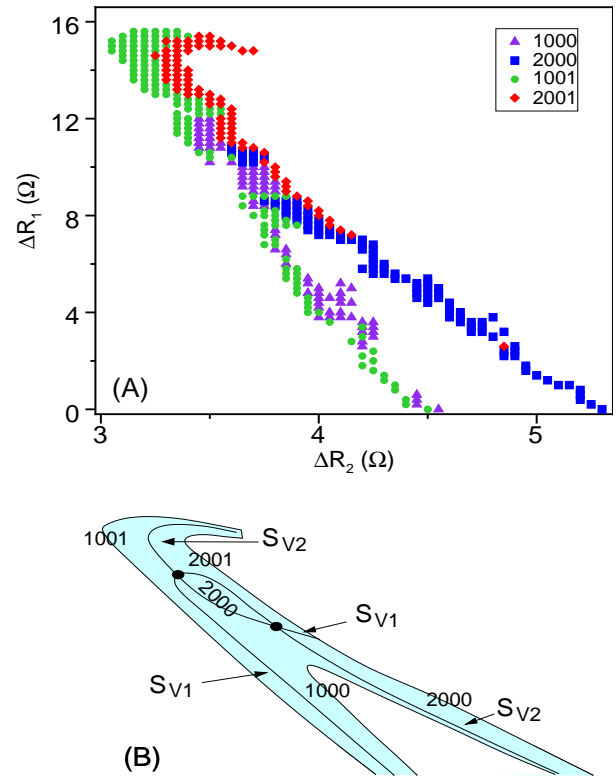}
\end{center}
\caption{[Color online] (A) The encoding of all the period-4 POs found in
the CPW. (B) Sketch of the critical lines ($S_{V1}$ and $S_{V2}$) structure
of the CPW, disregarding the existence of characteristic (iii) that causes
the appearance of structures as illustrated in Fig. \protect\ref{fig4}.}
\label{fig8}
\end{figure}


\begin{figure}[tbh]
\begin{center}
\includegraphics*[width=4.0cm, height=2.0cm,viewport= 0 0 200 150]{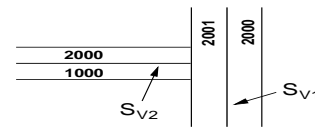}
\end{center}
\caption{Illustration of the structures that might appear in a CPW due to
characteristic (iii).}
\label{fig4}
\end{figure}

Concerning characteristic (i), for quadratic maps one should expect that 
\begin{equation}
\Delta \mathcal{P}(P) \propto e^{-\beta P}  \label{rene}
\end{equation}
\noindent as shown in Ref. \cite{hunt:1997}, with $\mathcal{P}$ being the
parameter interval length of a CPW, and $P$ the period of the lowest-period
periodic attractor. Also, from \cite{hunt:1997}, we have that $\beta \cong
2H_T$, where $H_T$ is the topological entropy or Lyapunov exponent of the
bordering chaotic region \cite{barreto:1997,murilo1}. But, in fact, for flows
such as the Chua's circuit containing Shilnikov's homoclinic orbits
\cite{comment_orbit}, a two-parameter analysis \cite{gaspard1984} performed in the
neighborhood of this orbit shows that it exists a countable (infinity) number
of CPWs that appear side-by-side in parameter space following the same
exponential scaling law that describe the appearance of the homoclinic
orbits. This exponential scaling law is of the form of Eq. (\ref{rene}), and as
shown in Ref. \cite{rene:2006}, $\beta=\pi\frac{\rho}{\omega}$, with $\rho$
and $\omega$ representing the real and imaginary parts of the eigenvalues of
the focus point associated with the homoclinic orbit responsible for the
generation of the many CPWs.

We estimate that for this experimental circuit $\beta \approxeq 2$, in
Eq. (\ref{rene}), for a parameter region in the vicinity of the observed
period-4 CPW. That means that in order to observe a higher period CPW, with
period $P_h=4^{2n}$, with $n \in \mathcal{N}$, associated with the observed
period-4 CPW, we should have a potentiometer with a resolution (step size) of $8 \Delta p
\exp^{(-2(P_h-4))}$, 8 being roughly an average width of the
period-4 CPW observed. So, in order to observe a period-8 CPW, we would need a
potentiometer with a resolution of about 0.14m$\Omega$, which is much smaller
than our experimental resolution. Numerical simulations realized in a similar
Chua's circuit, reported in Ref. \cite{baptista:1997}, show that CPWs with
attractors of period lower than 4 exist. However, their sizes are of the order
of 20 times smaller than a period-4 large CPW, similar to the one observed
experimentally. Therefore, for the resolution of our experiment, we do not
expect to find the many others numerically found CPWs, but only this "giant"
one.

To detect the existence of the critical curves, we search for transitions in
the symbolic sequence of the POs closer to the borders between the PW and
chaos. In box II, the PO encoded by ${1001}$ at the border \textbf{a} changes
its encoding to ${2001}$ at the border \textbf{b}. So, between these two
borders, there is a parameter $\Delta R_{2}$ for which at least one point of
the period-4 orbit is $V_{C_{1}}^{i}=V_{2}$. Thus, within these borders there
must exist a curve $S_{V2}$. In box III, the PO encoded by ${1001}$ (border
\textbf{a}) changes its encoding to ${1000}$ (border
\textbf{b}), indicating that within these borders there is a PO that visits the
critical point $V_{1}$. Thus, within these borders there must exist a curve
$S_{V1}$. In box III, the POs in both borders \textbf{c} and \textbf{d} are
encoded by the symbolic sequence ${2000}$, which suggests that within these two
borders there must exist either (or both) curves $S_{V1}$ or $S_{V2}$.

As we go from one side of the CPW to the other side by changing $\Delta R_{2}
$, for a fixed $\Delta R_{1}$, the points of the return map of the POs wander
along an imaginary smooth curve $\xi ^{\prime }$. This imaginary curve changes
its form smoothly, as we vary $\Delta R_{2}$. For a $\Delta R_{2}$ close to a
parameter where chaos is found (close to the borders {\bf a}, {\bf b}, {\bf c}
or {\bf d}), $\xi ^{\prime}$ resembles the return maps $\xi$ of the chaotic
attractors.  The curve $\xi ^{\prime}$ can be constructed using all the POs
observed in this CPW, for a constant $\Delta R_{1}$. Then, we estimate the
location of the critical points of $\xi ^{\prime }$, which provide us the
encoding for the period-4 POs within the CPW, in Fig. \ref{fig8}(A). The
curves $S_{V1}$ and $S_{V2}$ are located where two different colors (which 
describe the different encodings) meet. A curve $S_{V1}$ is the border line
between two regions representing different encodings. Either '1001' and
'1000', or '2001' and '2000'. A curve $S_{V2}$ is the border line between the
regions that encode either '1001' and '2001', or '1000' and '2000'. Note that
these curves cross transversally at least twice inside the windows, at the
points where the regions that encode the four different types of POs
meet. This is characteristic (ii) of a CPW \cite{murilo1}. It can be
understood by the way CPWs appear in the parameter space. The process can be
described as having a normal PW which contains two curves $S_{V1}$ and 
$S_{V2}$ that do not cross. One can imagine that both curves have a parabolic
shape appearing side by side. As one changes a parameter of the circuit, the
curve $S_{V2}$ approaches $S_{V1}$ crossing it in at least two points forming
a structure similar to the one shown in Fig. \ref{fig8}(B), a sketch of a
simplified version of what it could be really happening inside the CPW.
There, one sees that some regions in the parameter space that represent POs
with some encoding (e.g. '1001') do not border a region with some other
encoding ('2000'), except for the point where the curves $S_{V1}$ and $S_{V2}
$ cross. And when that happens (excluding the atypical case when the curves
are tangent), there has to be at least one more crossing inside the CPW, so
that the POs appear side by side other allowed POs. The rule is '1001' appears
aside '1000', which appears aside '2000', which appears aside '2001', which
appears aside '1001'.

Such a rule can be apparently violated due to characteristic (iii) which leads to
points where two or three different regions meet, as represented in
Fig. \ref{fig4}. But note that, in fact, the line $S_{V1}$ does not cross the
line $S_{V2}$, and thus, the rule that describes the crossing between these
lines is not violated. Internal noise and parameter fluctuations of the
circuit partially destroy the CPW. Adding the fact that we have limitations
in our parameter resolution, we do not expect to identify all these fine
details of the CPW, but rather a lower resolution picture, in which this rule
might be apparently violated.

\textbf{Acknowledgments } We thank the financial support of CNPq, FAPESP,
and discussions with R. O. Medrano-T concerning the relationship between
homoclinic orbits and the CPWs.


\begin{thebibliography}{99}
\bibitem{comment_PW} Since the parameter interval occupied by higher-period
attractors decreases in a power-law fashion with the period of the
attractor, we do not expect to observe the higher-period attractors of a PW,
in an experiment. In this work, we only consider the lowest-period periodic
attractor in a PW.
\bibitem{comment_orbit} A Shilnikov homoclinic orbit is an orbit homoclinic
to a saddle-focus point for which $|\rho/\lambda|<1$, where $\lambda$ is one
real eigenvalue and $\rho$ is the real part of the other two complex
conjugate eigenvalues.
\end{thebibliography}
\end{document}